\documentclass[fleqn,11pt,a4paper]{ClassHal}
\usepackage{microtype}
\usepackage[utf8]{inputenc}
\usepackage[T1]{fontenc}
\usepackage[american]{babel}
\usepackage[autostyle]{csquotes}
\usepackage{siunitx}
\usepackage{calc}
\usepackage{pgfplots}
\usepgfplotslibrary{groupplots}
\usepackage{xcolor}
\definecolor{ColorDoc}{cmyk}{.9,.5,0,.35}
\definecolor{ColorDoi}{cmyk}{.7,0,.3,.5}
\usepackage{soul}
\sethlcolor{yellow}
\soulregister\cite7
\soulregister\citep7
\soulregister\citet7
\soulregister\eqref7
\soulregister\ref7
\usepackage{amsmath, amssymb}
\usepackage{graphicx}
\usepackage{algorithm}
\usepackage{algpseudocode}
\usepackage{pifont}
\newcommand{\cmark}{\ding{51}}%
\newcommand{\xmark}{\ding{55}}%
\usepackage{tikz}
\usetikzlibrary{calc,patterns,decorations.pathmorphing,decorations.markings}
\usetikzlibrary{positioning,arrows}
\usepackage{subfigure}
\title{Latent-Lagrangian Neural Networks for Reduced Order Modeling of Non-autonomous Nonlinear Dynamical Systems
}
\author{Anand Kumar \textsc{Agrawal}\textsuperscript{1},Anders \textsc{Thorin}\textsuperscript{1}\\[0pt] {\scriptsize \textsuperscript{1} Université Paris-Saclay, CEA List, Palaiseau, France}}

\date{}

\begin{document}

\maketitle

\begin{abstract}
This work proposes a latent Lagrangian-based framework for reduced-order modelling of forced nonlinear dynamical systems. In contrast with conventional Lagrangian or Hamiltonian neural networks, our approach learns a set of latent coordinates sufficient to capture the dynamics conjointly with two neural networks for the latent kinetic and latent potential energies, and leverages force supervision to eliminate the need for an ODE solver during training. Consistency of physical laws in the latent space is ensured through the principle of virtual work. Results show that the model effectively learns the subtle dynamics induced by the system’s nonlinearity and non-convex potential energy, and generalizes well to unseen forces and initial conditions. These observations confirm the physical relevance of the proposed approach, and its interest for model reduction.
\end{abstract}



\section{Introduction}\label{sec:intro}

Nonlinear dynamical systems are ubiquitous in structural dynamics as well as in many other areas of science and engineering \cite{Strogatz2024, Emam2025, Rega2025}. Developing accurate models for such systems is essential for predicting their future behavior, improving performance, and, in some applications, enabling effective control. These systems have widely been modeled using physics-based approaches, typically in the form of differential equations derived from first principles. These governing equations are then solved using classical numerical methods to simulate system behavior. However, obtaining such equations requires accurate modeling of both the constitutive response and energy dissipation mechanisms, which often involves sophisticated experimental procedures. For many real-world complex systems, this level of characterization is difficult or even infeasible.

Data-driven modeling has emerged as a powerful alternative, leveraging machine learning techniques to infer system dynamics directly from observed data. Neural networks, in particular, have shown remarkable success in capturing intricate nonlinear relationships. However, conventional fully connected neural networks (FCNNs) typically model dynamics as memoryless black-box mappings between consecutive time steps (e.g., \( \mathbf{s}_t \mapsto \mathbf{s}_{t+1} \)), relying solely on input--output data pairs~\cite{choi2021data}. These models do not capture temporal dependencies beyond a single step and often fail to generalize beyond the training data~\cite{pan2018long}.

To overcome this issue, recurrent architectures such as Recurrent Neural Networks (RNNs) \cite{park2022recurrentneuralnetworksdynamical, mohajerin2018multisteppredictiondynamicsystems}, Long Short-Term Memory networks (LSTMs)~\cite{bonassi2020lstm}, and Gated Recurrent Units (GRUs)~\cite{bonassi2021stability} introduce an internal hidden state \( \mathbf{h}_t \) to capture temporal dependencies across multiple time steps. Usually these models define system evolution through discrete-time mappings of the form:
\[
\mathbf{h}_t = f_\theta(\mathbf{h}_{t-1}, \mathbf{s}_t), \quad \mathbf{s}_{t+1} = g_\theta(\mathbf{h}_t),
\]
 where \( \mathbf{s}_t \in \mathbb{R}^n \) is the observable system state, and \( \theta \) denotes the learnable parameters. While this formulation allows RNNs to model more complex temporal behavior than FCNNs, the dynamics are still learned as discrete transitions without any explicit model of the underlying continuous-time system.

In contrast, Neural Ordinary Differential Equations (Neural ODEs)~\cite{chen2018neural, NEURALODE2} model the system dynamics in continuous time via a parameterized vector field:
\[
\frac{d\mathbf{s}}{dt} = f_\theta(\mathbf{s}, t),
\]
allowing the state trajectory \( \mathbf{s}(t) \) to be computed via numerical integration. Neural ODEs provide a more natural framework for learning continuous-time dynamical systems, particularly in scenarios involving irregular temporal sampling.

However, since each of the aforementioned architectures, FCNNs, RNNs, or Neural ODEs, completely ignore the physical laws governing the system dynamics they often suffer from limited interpretability, poor generalization, and data inefficiency. 

Another possibility is to exploit the strengths of both: the ability of data-driven approaches to learn system dynamics solely from input–output data, without the need for accurate constitutive models or energy-dissipation mechanisms, and the ability of physics-based modeling approaches to be consistent with physical laws.

Recently, several attempts have been made to address the above-mentioned issues using neural networks which are constrained to satisfy some aspects of the physical behavior of the dynamical system either by modification of the loss function or by ingenious architectures. The Hamiltonian Neural Networks (HNNs) by Greydanus et al.~(2019) \cite{greydanus2019hamiltonian} is one of the early examples of embedding physical structure into deep learning models for dynamical systems. HNNs learn a scalar-valued Hamiltonian from which the vector field of the system is derived using Hamilton's equations. This formulation enforces the energy conservation as well as respects symplectic structure.
However, HNNs require canonical coordinates, which limits their practical applicability.
This limitation is addressed by Deep Lagrangian Networks (DeLaNs) \cite{lutter2019deep} and Lagrangian neural network \cite {cranmer2020lagrangian}, which do not require canonical coordinates. DeLaNs learn a fully connected neural network with the following outputs: (1) the off-diagonal and diagonal elements of the Cholesky factor of the configuration-dependent inertia matrix, and (2) the generalized gravitational force vector (i.e., the gradient of the potential energy). These outputs are used within the structure of the Euler–Lagrange equations and used in the loss function, thereby incorporating a strong inductive bias based on Lagrangian mechanics. While DeLaNs are effective, their reliance on a quadratic structure of kinetic energy restricts their generality.

On the other hand, Lagrangian Neural Networks (LNNs) \cite{cranmer2020lagrangian, LNNconstrained2025,tripura2024data} do not make any restriction on the structure of the kinetic energy and learn a general Lagrangian using a parameterized FCNN which is then used to form the vector field  by enforcing the output of the network to satisfy the Euler-Lagrange equation. However, the monolithic modeling of the Lagrangian limits modular control over the architectures of potential and kinetic energies. Also, the results were demonstrated only on autonomous systems without the presence of forcing. 

Although these works (HNNs, DeLaNs, LNNs etc.) enhanced the interpretability and the generalization capacity of the neural models for the dynamical systems, they were not designed specifically towards model reduction and high-dimensional nonlinear dynamical systems. However, a recent work by Sharma et al.~\cite{sharma2024lagrangian} focuses on dimensionality reduction by making use the ideas in LNN: it projects data onto a fixed POD subspace, identifies a linear Lagrangian ROM, and then augments it with structure-preserving ML terms to capture the missing nonlinear operators. Key limitations include dependency on a prescribed linear basis, step-size sensitivity from finite-difference derivative estimation, limited treatment of external forcing, and restricted architectural control over the kinetic and potential energy network-structure. 

Latent-Energy-Based Neural Networks (LEBNNs) by Pottier et al.~\cite{pottier2025latent} attempted to incorporate physics similar to the case of LNNs and HNNs into the neural network model while also performing model order reduction by learning two autoencoders along with the potential energy network. Here, they  addressed the statics of hyperelastic systems with non-convex strain energy using neural networks. The capability to reduce the model order of a mechanics problem using neural networks while being physically consistent makes this work quite useful for high-dimensional statics problems. Key innovations in their work include the use of double autoencoders for compressing displacements and forces into low-dimensional latent spaces, learning a latent energy function from which forces are computed as gradients, and dealing with multiple equilibrium states through the minimization of latent energy.

LEBNNs offer good performance and interpretability in static scenarios but are not formulated to handle time-dependent dynamics as they omit kinetic energy. Moreover, by learning latent forces separately from displacements using two independent autoencoders, they break the fundamental energy–force relationship, making the forces non-variational. As a result, the virtual work consistency may not hold between the latent and original coordinate spaces.


This work aims to address the aforementioned limitations by developing a Latent-Lagrangian Neural Network (L-LNN), that can be seen as a variant of a Lagrangian NN \cite{cranmer2020lagrangian, LNNconstrained2025} modified so that the Lagrangian is learned in a low-dimensional latent space. Additionally, the training is performed in a manner that allows modular control over the architectures for kinetic and potential energy networks which allows the additional knowledge about the nature of these energies to be used in the design of their architectures. Apart from this, our networks are trained using force supervision which avoids using the ODE solver inside the training loop. It can also be seen as an extension of Latent-Energy Based NN (LEBNN) \cite{pottier2025latent} to dynamics where the consistency between the force fields in the latent space and original space is ensured by balancing the virtual work done by these force fields.  Therefore, this approach comprehensively enforces the governing mathematical model while being cost effective. 

The methodology involves simultaneous training of three fully connected neural networks (i.e., the auto-encoder, the potential energy, and the kinetic energy) using a single composite loss function formed by embedding the energy and dissipation-aware terms into the loss function of the neural networks through mean squared error in the true and predicted force as well as the reconstruction loss of the auto-encoder. Specifically, the neural networks are designed to satisfy the structure of the physics-based model through position-dependent mass and stiffness matrices which are written as Hessian of parameterized neural networks representing latent kinetic and potential energies, respectively. The damping matrix is modeled via a position-dependent linear combination of the mass and stiffness matrices (position-dependent Rayleigh damping). The above-mentioned approach ensures good predictive capability, as well as computational and data efficiency.

{\footnotesize%
\begin{table}[h!]
\caption{Comparison of learning-based physics models. $T$ denotes kinetic energy, $V$ potential energy. }
\centering
\renewcommand{\arraystretch}{1.2}
\setlength{\tabcolsep}{2.9 pt}
\begin{tabular}{p{3.5cm} p{1.2cm} p{1.2cm} p{1.1cm} p{1.1cm} p{1.1cm} p{1.2cm} p{1.6cm} p{1.6cm}}
\hline
 Feature & Neural net & Neural ODE & HNN & DeLaN & LNN & LEBNN & LOpInf-SpM~\cite{sharma2024lagrangian}& Latent-LNN (ours) \\
\hline
Can model dynamical systems    & \cmark & \cmark & \cmark & \cmark & \cmark & \xmark &\cmark& \cmark \\

Learns differential equations  & \xmark     & \cmark & \cmark & \cmark & \cmark & \xmark & \cmark& \cmark \\
Learns exact conservation laws & \xmark     & \xmark     & \cmark & \cmark & \cmark & \cmark$^*$& \cmark & \cmark \\
Learns from arbitrary coords.  & \cmark & \cmark & \xmark     & \cmark     & \cmark & \cmark & \cmark & \cmark \\
Learns arbitrary Lagrangians   & \xmark     & \xmark     & \xmark     & \xmark & \cmark & no $T$& \cmark & \cmark \\
Separate control over architectures of $T$, $V$ & \xmark     & \xmark     & \xmark     & \xmark & \xmark & no $T$ & \xmark & \cmark \\
Model Order Reduction         & \xmark     & \xmark     & \xmark     & \xmark & \xmark & \cmark & \cmark & \cmark \\
Joint learning of latent space and $T$-$V$  & \xmark     & \xmark     & \xmark     & \xmark & \xmark & no $T$ & \xmark & \cmark \\
\hline
$^*$Static energy conservation only.
\end{tabular}

\label{tab:model_comparison}
\end{table}
}%

We validate the effectiveness of our approach through numerical experiments on nonlinear damped dynamical systems. 
This work contributes to the advancement of data-driven dynamical system modeling by incorporating physical principles, thereby improving both interpretability and generalization capabilities of the models. Table~\ref{tab:model_comparison} provides a summary of the comparison of the state-of-the-art approaches with the approach developed in work. The key contributions of this work can be summarized as follows:
\begin{enumerate}
  \item We use an \textbf{autoencoder} to learn latent coordinates $z \in \mathbb{R}^d$ from generalized coordinates $q$.Subsequently, obtain the latent velocities and accelerations from the latent coordinates by using derivative based relationship between them. 
  
  \item \textbf{Potential} $V(z)$ and \textbf{kinetic energy} $T(z, \dot{z})$ are modeled via separate neural networks, this provides modular control over the network architectures for the potential and kinetic energies, as opposed to using a monolithic network for learning the Lagrangian.
  \item In the proposed approach, the variational structure is preserved in the relationship between the forces in the latest space and the forces in the original space by enforcing  \textbf{the virtual work consistency}  which ensures that forces are derived from energy gradients and respect the energy -force duality under differentiation. In contrast, LEBNN learns latent forces separately, breaking the energy–force relationship and lacking a principled projection to physical space. 
  
  \item The model is trained using force supervision as opposed to the super-vision based on the trajectory data which helps in avoiding the use of an ODE-solver inside the training loop
 
  \item Modular and interpretable energy components
\end{enumerate}

The remainder of this paper is organized as follows: Section 2 presents the developed methodology. Section 3 details the results obtained from performance validation on two benchmark nonlinear dynamical systems. Section 4 concludes the paper and provides future directions.

\section{Proposed method:  Latent-Lagrangian Neural Networks}
\label{sec:LLNN}

This section details the latent Lagrangian-based methodology proposed in this work for the reduced-order modeling of the non-autonomous nonlinear dynamical systems. The methodology involves simultaneous training of three neural networks: an auto-encoder for the dimensionality reduction, one for the potential energy network and another for the kinetic energy network. The loss function encodes the structure of the governing Lagrangian equation by incorporating the discrepancy between the predicted and the true force values as well as a reconstruction loss of the autoencoder. The detailed discussion of the formulation of the training strategy is presented in the following subsections.

\subsection{Latent-Space Kinematics via Autoencoder}
\label{subsec:latent-kin}

Let \( \mathbf{q} \in \mathbb{R}^n \) be the vector of generalized coordinates of the system. The encoder part  $\mathcal{E}$ of the autoencoder maps these vectors to \( \mathbf{z} \in \mathbb{R}^m \) where $m\leq n$:
\begin{equation}
\mathbf{z} = \mathcal{E}(\mathbf{q}).
\label{eq:enc}
\end{equation}
The decoder part $\mathcal{D}$ maps $\mathbf{z}$ to a reconstruction $\hat{\mathbf{q}}$ of the generalized coordinates:
\begin{equation}
\hat{\mathbf{q}} = \mathcal{D}(\mathbf{z}),
\label{eq:dec}
\end{equation}
Once the autoencoder is trained, $\hat{\mathbf{q}}$ should be as close as $\mathbf{q}$ as possible. The discrepancy between these two is the \emph{reconstruction error}, defined precisely later.

The transformation of vectors of generalized velocity $\dot{\mathbf{q}}$ and acceleration $\ddot{\mathbf{q}}$ that is consistent with the coordinate transformation performed in \eqref{eq:enc} can be achieved by the application of chain rule as given in \eqref{eq:vel} and \eqref{eq:acc} respectively.
The latent velocity vector is obtained using the chain rule as follows:

\begin{equation}
\dot{\mathbf{z}} = \mathbf{J}(\mathbf{q}) \dot{\mathbf{q}}
\label{eq:vel}
\end{equation}
where, the Jacobian of the encoder is defined as:`
\begin{equation}\label{eq:Jenc}
\mathbf{J}(\mathbf{q}) = \frac{\partial \mathbf{z}}{\partial \mathbf{q}} \in \mathbb{R}^{m \times n}
\end{equation}
To compute the latent acceleration vector, differentiate with respect to time:

\begin{equation}
\ddot{\mathbf{z}} = \frac{d}{dt} (\mathbf{J}(\mathbf{q}) \dot{\mathbf{q}}) = \mathbf{J}(\mathbf{q}) \ddot{\mathbf{q}} + \frac{d\mathbf{J}}{dt} \dot{\mathbf{q}}
\label{eq:acc}
\end{equation}
where, \( \mathbf{J} \) depends on $t$ implicitly via \( \mathbf{q}(t) \), therefore, the time derivative is computed as given by:
\begin{equation}
\left[ \frac{d \mathbf{J}}{dt} \right]_{ij} = \sum_{k=1}^{n} \frac{\partial^2 z_i}{\partial q_j \partial q_k} \dot{q}_k
\label{eq:djdt}
\end{equation}
Here, $\left[\tfrac{d\mathbf{J}}{dt}\right]_{ij}$ denotes the $(i,j)$-th entry of the time derivative of the Jacobian matrix $\tfrac{d\mathbf{J}}{dt}$, i.e., the rate of change of $\tfrac{\partial z_i}{\partial q_j}$. 
Thus, the full latent acceleration becomes:

\begin{equation}
[\ddot{\mathbf{z}}]_i = \sum_{j=1}^{n} \frac{\partial z_i}{\partial q_j} \ddot{q}_j + \sum_{j=1}^{n} \sum_{k=1}^{n} \frac{\partial^2 z_i}{\partial q_j \partial q_k} \dot{q}_j \dot{q}_k
\label{eq:acc1}
\end{equation}
where, $[\ddot{\mathbf{z}}]_i$ denotes the $i$-th component of the latent acceleration vector $\ddot{\mathbf{z}}$.
Equation \eqref{eq:acc1} splits the acceleration into an inertial term and a geometric term, also referred to as linear and non-linear acceleration terms in the literature.

\subsection{Latent-Space Lagrangian Dynamics via Potential and Kinetic Energy Networks}

The latent descriptions of the generalized coordinates, $\mathbf{z}$, and velocities, $\dot{\mathbf{z}}$ obtained in Subsection~\ref{subsec:latent-kin} can be utilized to define a Lagrangian in the latent space as: 

\begin{equation}
\mathcal{L}(\mathbf{z}, \dot{\mathbf{z}}) = T(\mathbf{z}, \dot{\mathbf{z}}) - V(\mathbf{z})
\end{equation}
where, $T(\mathbf{z}, \dot{\mathbf{z}})$ denotes the kinetic energy in the latent space and can be modeled using either of two neural network architectures the parameters of which are denoted generically by $\theta$:
\begin{itemize}
    \item A structured neural network that outputs a kinetic energy with coordinate-dependent symmetric mass matrix, and quadratic form in generalized velocity as \( \frac{1}{2} \dot{\mathbf{z}}^\top \mathbf{W}_\theta(\mathbf{z}) \dot{\mathbf{z}} \). This mimics classical mechanical formulations and maintains interpretability. 
    \item An unconstrained neural network \( \mathcal{T}_\theta(\mathbf{z}, \dot{\mathbf{z}}) \) that directly maps inputs to scalar kinetic energy, offering more modeling flexibility.
\end{itemize}
Both are trainable and differentiable with respect to their parameters, and either can be used in the force derivation depending on the problem at hand.

Potential energy \( V(\mathbf{z}) \) is learned using a separate fully connected neural network.
These neural networks for kinetic and potential energy are used together to derive the latent dynamics using the Euler-Lagrange equation:

\begin{equation}
\frac{d}{dt} \left( \frac{\partial \mathcal{L}}{\partial \dot{\mathbf{z}}} \right) - \frac{\partial \mathcal{L}}{\partial \mathbf{z}} = \mathbf{f}_{\text{latent}}
\end{equation}

Expanding the terms:

\begin{equation}
\frac{d}{dt} \left( \frac{\partial \mathcal{L}}{\partial \dot{\mathbf{z}}} \right) = \frac{\partial^2 T}{\partial \dot{\mathbf{z}}^2} \ddot{\mathbf{z}} + \frac{\partial^2 T}{\partial \mathbf{z} \partial \dot{\mathbf{z}}} \dot{\mathbf{z}}
\end{equation}

\begin{equation}
\frac{\partial \mathcal{L}}{\partial \mathbf{z}} = \frac{\partial T}{\partial \mathbf{z}} - \frac{\partial V}{\partial \mathbf{z}}      
\end{equation}

The latent-space force is then:

\begin{equation}
\mathbf{f}_{\text{latent}} = \underbrace{\frac{\partial^2 T}{\partial \dot{\mathbf{z}}^2} \ddot{\mathbf{z}}}_{\text{inertial}} + \underbrace{\frac{\partial^2 T}{\partial \mathbf{z} \partial \dot{\mathbf{z}}} \dot{\mathbf{z}}}_{\text{mixed}} - \underbrace{\frac{\partial T}{\partial \mathbf{z}}}_{\text{non-conservative}} + \underbrace{\frac{\partial V}{\partial \mathbf{z}}}_{\text{potential}}
\end{equation}

\subsection{Incorporating Damping and Final Force Expression}
Damping is often introduced in structural mechanics as a linear combination of the mass and damping matrices, a method known as Rayleigh damping. In this work, a similar approach was used, however including a position-dependency in the mass and damping matrices, corresponding to the Hessian matrices of kinetic and potential energies. The weights of the linear combination \( \alpha, \beta \) are learnable scalars, so that the final force becomes:
\begin{equation}
\mathbf{f}_z \approx \mathbf{H}_T \ddot{\mathbf{z}} + (\alpha \mathbf{H}_T + \beta \mathbf{H}_V) \dot{\mathbf{z}} + \frac{\partial V}{\partial \mathbf{z}} + \frac{\partial^2 T}{\partial \dot{\mathbf{z}} \partial \mathbf{z}} \dot{\mathbf{z}} - \frac{\partial T}{\partial \mathbf{z}}
\label{eq:damp_incor}
\end{equation}
with the Hessians matrices \( \mathbf{H}_T = \frac{\partial^2 T}{\partial \dot{\mathbf{z}}^2} \) and \( \mathbf{H}_V = \frac{\partial^2 V}{\partial \mathbf{z}^2} \).

\subsection{Force relations via virtual work consistency}


The predicted latent-space force vector $\mathbf{f}_z \in \mathbb{R}^m$, obtained from the Euler--Lagrange equation in the latent space, must be projected back to the original coordinate space $\mathbf{q} \in \mathbb{R}^n$ to compare with the known physical force $\mathbf{f}_{\text{true}} \in \mathbb{R}^n$.
In order to be consistent with the principle of \emph{virtual work}, this projection is achieved through the transpose of the encoder Jacobian (see Eq.~\eqref{eq:Jenc}:
\begin{equation}
\hat{\mathbf{f}}_q = \mathbf{J}^\top \mathbf{f}_z
\label{eq:force}
\end{equation}
Under a virtual displacement $\delta \mathbf{q}$, the corresponding displacement in latent space is $\delta \mathbf{z} = \mathbf{J} \delta \mathbf{q}$. The virtual work done in both spaces should be equal:
\begin{equation}
\mathbf{f}_z^\top \delta \mathbf{z} = \hat{\mathbf{f}}_q^\top \delta \mathbf{q}
\end{equation}
Substituting $\delta \mathbf{z} = \mathbf{J} \delta \mathbf{q}$, we get:
\begin{equation}
\mathbf{f}_z^\top \mathbf{J} \delta \mathbf{q} = \hat{\mathbf{f}}_q^\top \delta \mathbf{q} \quad \Rightarrow \quad \hat{\mathbf{f}}_q = \mathbf{J}^\top \mathbf{f}_z
\end{equation}
This ensures that the predicted force in the original coordinates is consistent with the variational principle and is physically meaningful.

\subsection{Formulation of the Loss Function}
All networks are trained together by minimizing the loss function that is defined in this section. The loss function is formulated by utilizing
the reconstructed generalized coordinate vector $\hat{q}$ obtained from the decoder $\mathcal{D}$ in equation~\eqref{eq:dec} and the predicted force vector $\hat{\mathbf{f}}_q$  in equation \eqref{eq:force} as defined in equation~\eqref{eq:loss}. This final loss function is minimized for the simultaneous learning of the parameters of all the networks as well as the parameters (\( \alpha, \beta \)) used to incorporate damping in equation~\eqref{eq:damp_incor}.

\begin{equation}
\mathcal{L}_{\text{total}} = \mathcal{L}_{\text{recon}} + \mathcal{L}_{\text{force}} = 
\| \hat{\mathbf{q}} - \mathbf{q} \|^2 + \lambda\| \hat{\mathbf{f}}_q - \mathbf{f}_{\text{true}} \|^2
\label{eq:loss}
\end{equation}
where $\mathcal{L}_{\text{recon}}$ denotes the reconstruction loss, $\mathcal{L}_{\text{force}}$ represents the force loss and the {$\lambda$ is a weighting parameter that controls the relative importance of the reconstruction loss and the force loss.

A similar formulation of the loss function, wherein the force-matching term employed in the present work is substituted with a weighted discrepancy in the lift coefficient, has been adopted in prior studies on observable-augmented autoencoders \cite{fukami2023grasping,fukami2025compact}. Specifically, the primary reconstruction term in \cite{fukami2023grasping} corresponds to the mean squared error on the vorticity field, whereas in \cite{fukami2025compact}, it pertains to the pressure field; in both cases, the auxiliary lift loss is scaled by a factor selected via L-curve analysis to balance the importance to the field accuracy with aerodynamic interpretability.} 

\begin{figure}[h!]
    \centering
    \includegraphics[width=\textwidth]{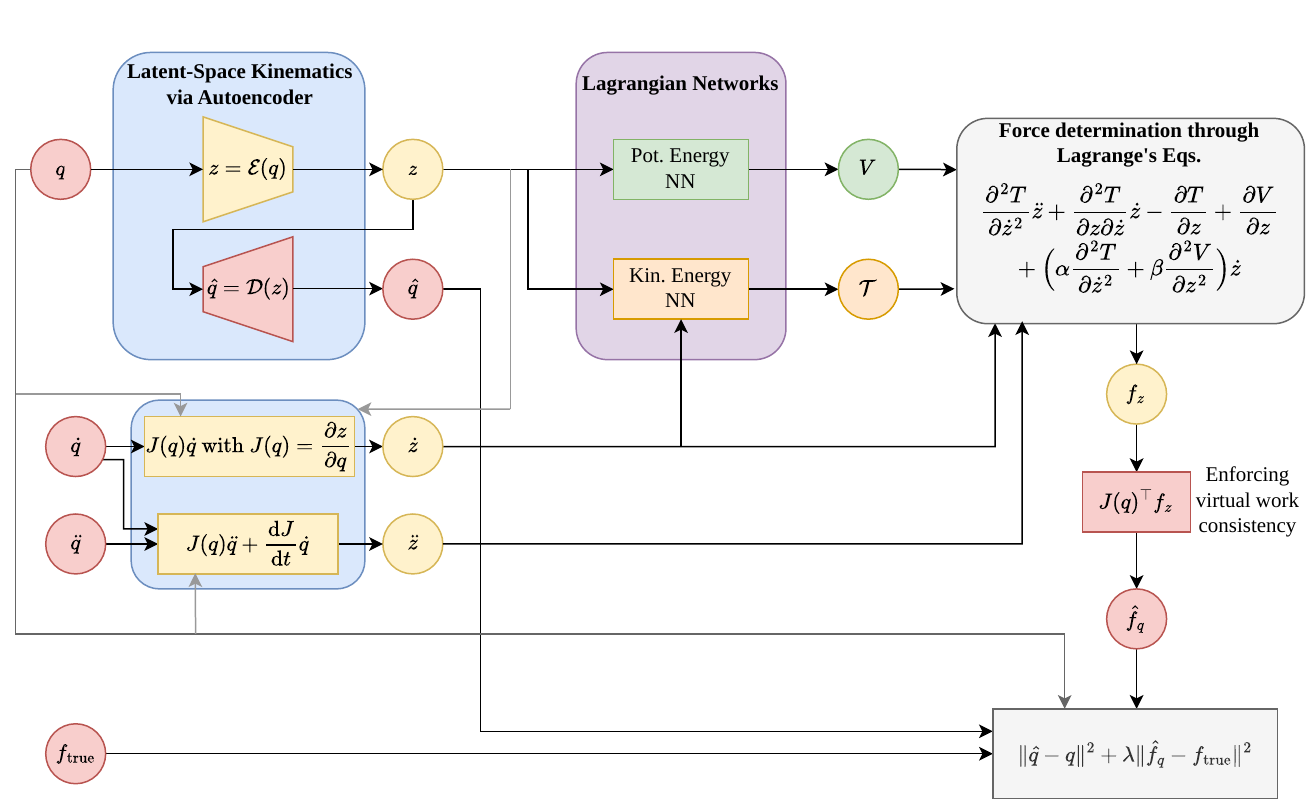}
    \caption{Graphical representation of Latent-LNN}
    \label{fig:meth}
\end{figure}

\subsection{Graphical and algorithmic summaries}
The details of the computer implementation of the developed methodology is provided in the form of a pseudo code in 
Algorithm~\ref{alg:llnn}. Also, a schematic of the forward pass of the composite neural network training is shown in Figure~\ref{fig:meth}.

\begin{algorithm}[H]
\caption{Latent Energy-Based Neural Network Training with Force Supervision}\label{alg:llnn}
\begin{algorithmic}[1]
\Require Dataset $\mathcal{D} = \{q, \dot{q}, \ddot{q}, f_{\text{true}}\}$, networks: $AE (q)$, $T(z, \dot{z})$, $V(z)$, scalars $\alpha$, $\beta$, $\lambda$, learning rate $\eta$, number of epochs $N$
\For{each epoch $= 1$ to $N$}
    \For{each mini-batch $\{q, \dot{q}, \ddot{q}, f_{\text{true}}\} \in \mathcal{D}$}
        \State Encode $z \gets \text{Encoder}(q)$
        \State Decode $\hat{q} \gets \text{Decoder}(z)$
        \State Compute Jacobian $J = \partial z / \partial q$ and Hessian $H = \partial^2 z / \partial q^2$
        \State Compute latent velocity: $\dot{z} \gets J \dot{q}$
        \State Compute latent acceleration: $\ddot{z} \gets J \ddot{q} + \sum_{j,k} \frac{\partial^2 z}{\partial q_j \partial q_k} \dot{q}_j \dot{q}_k$
        \State Evaluate energies: $T \gets T(z, \dot{z})$, $V \gets V(z)$
        \State Compute gradients:
        \begin{itemize}
            \item $H_T \gets \partial^2 T/\partial \dot{z}^2$
            \item $H_V \gets \partial^2 V/\partial z^2$
            \item $\nabla_z V \gets \partial V/\partial z$
            \item $\partial^2 T/\partial \dot{z} \partial z$, $\partial T/\partial z$
        \end{itemize}
        \State Compute latent-space force:
        \[
        f_z \gets H_T \ddot{z} + (\alpha H_T + \beta H_V) \dot{z} + \nabla_z V + \frac{\partial^2 T}{\partial \dot{z} \partial z} \dot{z} - \frac{\partial T}{\partial z}
        \]
        \State Compute encoder Jacobian: $J_{\text{enc}} \gets \partial z / \partial q$
        \State Project to physical coordinates: $\hat{f}_q \gets J_{\text{enc}}^\top f_z$

        \State Compute total loss: $\mathcal{L} \gets \lambda \|\hat{f}_q - f_{\text{true}}\|^2 + \|\hat{q} - q\|^2$
        \State Backpropagate and update parameters of $AE$, $T$, $V$, $\alpha$, $\beta$ using optimizer
    \EndFor
\EndFor
\State \Return Trained networks $AE$, $T$, $V$, and scalars $\alpha$, $\beta$
\end{algorithmic}
\end{algorithm}

\section{Validation methodology}\label{sec:demos}

The present section outlines the validation approach adopted to demonstrate the performance and relevance of the proposed Latent-Lagrangian Neural Network framework for model reduction. To this end, two different dynamical systems are used. The validation process begins with generating the training data from each of these dynamical systems. 

\begin{algorithm}[H]
\caption{Validation workflow after training the networks} \label{alg:validation}
\begin{algorithmic}[1]

\Require Trained models (from Algorithm~\ref{alg:llnn}): Autoencoder $AE$, 
Kinetic Energy Network $T(z,\dot{z})$, 
Potential Energy Network $V(z)$; 
parameters: damping scalars $(\alpha,\beta)$, forcing $f_q(t)$; 
initial conditions $q(0), \dot{q}(0)$.

\textbf{Latent-Space Simulation:}
\State Encode $q(0)$ into latent coordinates: $z(0) = AE.\text{encoder}(q(0))$.
\State Map initial velocity: $\dot{z}(0) = J(q(0))\dot{q}(0)$, with $J=\partial z / \partial q$.
\State Initialize latent state: $y(0) = [z(0), \dot{z}(0)]$.

\For{each timestep $t$}
  \State Compute latent mass matrix $H_T = \dfrac{\partial^2 T}{\partial \dot{z}^2}$.
  \State Compute stiffness matrix $H_V = \dfrac{\partial^2 V}{\partial z^2}$ and elastic force $\nabla_z V$.
  \State Compute mixed term $\dfrac{\partial^2 T}{\partial \dot{z}\partial z}\dot{z}$ and $-\dfrac{\partial T}{\partial z}$.
  \State Form damping matrix: $C = \alpha H_T + \beta H_V$.
  \State Map external force $f_q(t)$ from physical to latent space: 
  \State \hspace{1em} $f_z(t) = J_{\text{dec}}^\top f_q(t)$, with $J_{\text{dec}} = \dfrac{\partial \hat{q}}{\partial z}$.
  \State Solve latent ODE:
  \State \hspace{1em} $H_T(z,\dot{z}) \ddot{z} + C(z,\dot{z})\dot{z} + \nabla_z V(z)$
  \State \hspace{3em} $+ \Big(\dfrac{\partial^2 T}{\partial \dot{z}\partial z}\Big)\dot{z} - \dfrac{\partial T}{\partial z} = f_z(t)$.
  \State Update latent state $[z(t), \dot{z}(t)]$ using ODE solver.
\EndFor

\textbf{Decode trajectories back to original space:}
\State \hspace{1em} $q(t) = AE.\text{decoder}(z(t))$, 
$\dot{q}(t) = J(z(t))\dot{z}(t)$.

\textbf{True-System Simulation:}
\State Define nonlinear spring–mass model with damping and forcing in $(q,\dot{q})$.
\State Solve ground-truth system with \texttt{solve\_ivp} to obtain $q^{\text{true}}(t), \dot{q}^{\text{true}}(t)$.

\textbf{Validation:}
\State Compute relative errors:
\State \hspace{1em} $e_{q_i} = \dfrac{\|q_i^{\text{true}} - q_i^{NN}\|}{\|q_i^{\text{true}}\|}$, \;
$e_{\dot{q}_i} = \dfrac{\|\dot{q}_i^{\text{true}} - \dot{q}_i^{NN}\|}{\|\dot{q}_i^{\text{true}}\|}$.
\State Plot NN response vs.\ true response, error histories, and energy contours.

\end{algorithmic}
\end{algorithm}

These training data are then used to train the autoencoders and the latent-space energy networks using Algorithm~\ref{alg:llnn} for the corresponding system. After which, the learned networks are evaluated by following the approach summarized in Algorithm~\ref{alg:validation}. This involves obtaining the equations of motion (or the vector field) using the learned energy networks and comparing the decoded response (generalized coordinates, generalized velocity) obtained from the learned system with that of the true system of equations. To further strengthen the validation process, we have also made comparison between the learned and true topologies of the kinetic and potential energies. Further details of the two nonlinear dynamical systems used for validation, for the generation of the training data, and the details for the training are provided in subsequent subsections.

\subsection{Nonlinear systems used for validation}\label{sec:test_cases}
Both the aforementioned dynamical systems that are used for the purpose of validating the proposed method are detailed here. The first system is a two-degree-of-freedom (2-DOF) system with nonconvex potential energy, as shown in Figure~\ref{fig:2dof}. Despite its low-dimensional configuration space, the presence of large deflections and the nonconvexity of its potential energy give rise to a rich dynamics that exhibits nonlinearity and the existence of stable and unstable equilibria. Apart from this, its low dimensionality also allows the visual comparison between the learned and the true topologies of the kinetic and potential energy. The same system without the presence of dynamic behaviour and damping was used to assess the performance and relevance of Latent-Energy-Based NNs~\cite{pottier2025latent}, which were proposed for static problems as mentioned in section~\ref{sec:intro}. To make the description unambiguous, the mathematical model for this system has been presented in the appendix~\ref{app:2DOF}.

\begin{figure}[h]
    \centering
        \begin{tikzpicture}[scale=0.9]
    \tikzstyle{spring}=[thick,decorate,decoration={zigzag,pre length=0.3cm,post length=0.3cm,segment length=9}]
     \tikzstyle{spring2}=[thick,decorate,decoration={zigzag,pre length=0.3cm,post length=0.3cm,segment length=6}]

    \tikzstyle{ground}=[fill,pattern=north east lines,draw=none,minimum width=0.75cm,minimum height=0.3cm,inner sep=0pt,outer sep=0pt]

    \node (groundL) [anchor=east] at (0,0) {};
    \begin{scope}
    \clip (-0.3,-0.5) rectangle (0,0.5);
    \foreach \i in {-1.5,-1.2,...,1.5}
   {
        \draw (-0.3,\i) --++(0.3,-.5);
    }
    \end{scope}
    \draw[thick] (0,-0.5) -- ++(0,1);
   
        \begin{scope}
    \clip (1.5,-2) rectangle (2.5,-2.3);
    \foreach \i in {1,1.1,...,3}
    {
        \draw (\i,-2) --++(0.3,-1);
    }
    \end{scope}
    \draw[thick] (1.5,-2) -- ++(1,0);

    \node (groundB) [anchor=north] at (2,-2) {};
    
    \draw[spring] (0,0) -- (2.5,.7) node[midway,above] {$k_x$};
    \draw[spring] (2,-2) -- (2.5,.7) node[midway,right] {$k_y$};

    \fill (2.5,.7) circle (5pt) node[above,xshift=4mm] {$m$};
    
    \draw[thin,opacity=0.5,spring2] (0,0) -- (2.,0);
    \draw[thin,opacity=0.5,spring2] (2,-2) -- (2,0);

    \fill[gray] (2.0,0) circle (5pt);
    \draw[blue,xshift=10pt,->,>=latex] (2.5,0) -- (2.5,.7) node[right,midway] {${q_2}$};
    \draw[very thin] (2.5,0.7) -- (2.9,0.7);
    \draw[very thin] (2.5,0.7) -- (2.5,1.1);
    \draw[blue,yshift=10pt,->,>=latex] (2.,0.7) -- (2.5,.7) node[above,midway] {${q_1}$};
        \draw[thin,dashed] (0,0) -- (3,0);
    \draw[thin,dashed] (2,-2) -- (2,1);
    \draw[thick,->,>=latex,red] (2.5,0.7) -- ++(0.6,.75) node[right] {$\vec{f}_\text{true}$};

    \end{tikzpicture}
    \caption{2-DOF system used to assess and compare the proposed architecture. Springs are linear, but the geometry is nonlinear (large displacements). From~\cite{pottier2025latent}.}
    \label{fig:2dof}
\end{figure}
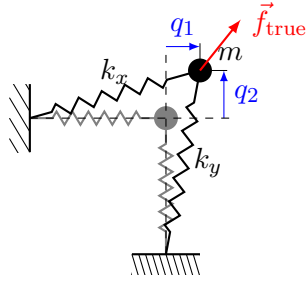
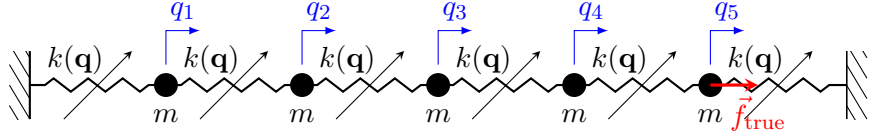
\begin{figure}[h]
    \centering
        \begin{tikzpicture}[scale=0.9]
    \tikzstyle{spring}=[thick,decorate,decoration={zigzag,pre length=0.2cm,post length=0.2cm,segment length=13}]
   
    \tikzstyle{ground}=[fill,pattern=north east lines,draw=none,minimum width=0.75cm,minimum height=0.3cm,inner sep=0pt,outer sep=0pt]
    
    \node (groundL) [anchor=east] at (0,0) {};
    \begin{scope}
    \clip (-0.3,-0.5) rectangle (0,0.5);
    \foreach \i in {-1.5,-1.2,...,1.5}
   {
        \draw (-0.3,\i) --++(0.3,-.5);
    }
    \end{scope}
    \draw[thick] (0,-0.5) -- ++(0,1);
   
   \begin{scope}[shift={(12.3,0)}]
    \clip (-0.3,-0.5) rectangle (0.5,0.5);
    \foreach \i in {-1.5,-1.2,...,1.5}
   {
        \draw (-0.3,\i) --++(0.3,-.5);
    }
    \draw[thick] (-0.3,-0.5) -- ++(0,1);
    \end{scope}

    \foreach \i in {1,2,...,5}
{
    \draw[spring] ($ ({2*(\i - 1)}, 0) $) -- ($ ({2*\i}, 0) $) 
        node[midway,above,xshift=-3mm] {$k(\mathbf{q})$};
    \fill (2*\i,0) circle (5pt) node[yshift=-4mm] {$m$};
   \draw[blue,->,>=latex] ($ ({2*\i},0.3)$) -- ++(0,.5) -- ++(.5,0) node[above,midway] {$q_{\i}$};
   \draw[->,>=stealth,thin] ($ ({2*(\i - 1) +.5}, -0.5) $) -- ($ ({2*\i -0.5}, 0.5) $) ;
}
\draw[spring] (10, 0) -- (12, 0) node[midway,above,xshift=-3mm] {$k(\mathbf{q})$};
\draw[->,>=stealth,thin] ($ ({2*(5) +.5}, -0.5) $) -- ($ ({2*6 -0.5}, 0.5) $);
   
   \draw[->,>=stealth,very thick, red] ($ ({2*(5) }, 0) $) -- ++(.7,0) node[below,red] {$\vec{f}_\textup{true}$}  ;
    \end{tikzpicture}
        \caption{Schematic of the 5-DOF nonlinear system. The last mass is subject to an external load. Springs have a nonlinear behaviour (cubic nonlinearity). }
    \label{fig:5dof}
    \end{figure}

The second system is a 5-DOF dynamical system with Duffing-like nonlinearities, depicted in Figure~\ref{fig:5dof}. Its potential energy includes a non-quadratic term:
\begin{equation}
V = \tfrac{1}{2}\,\mathbf{q}^T \mathbf{K} \mathbf{q} + \frac{\alpha_*}{4}\left(q_1^4 + q_2^4 + q_3^4 + q_4^4 + q_5^4\right).
\label{eq:PE}
\end{equation}In contrast to the low-dimensional 2-DOF case, this system provides a higher-dimensional configuration space and offers the ability to further assess the scalability and robustness of the proposed method when applied to a more complex multi-degree-of-freedom nonlinear system, in particular for model reduction, by choosing a latent space dimension smaller than the number of DOFs. The mathematical description for this system has been presented in the appendix~\ref{app:5DOF}.

\subsection{Data-generation and training}\label{sec:data}
For both systems, the training data contains several time series for displacement, velocity, acceleration, and force. The data were collected by numerically simulating the equations of motion for these systems, which are provided in appendix~\ref{app:2DOF} and \ref{app:5DOF}. Further details of the training data and the system parameters used during simulation to generate it are provided below. 

For generation of the dataset for the 2-DOF case, the parameters of the model in appendix~\ref{app:2DOF} were set as
$m = 1$, $k_x = k_y = 1$, $c_x = c_y = 0.1$,
with initial conditions
$q_1(0) = q_2(0) = \dot{q}_1(0) = \dot{q}_2(0) = 0$,
and external excitations
$F_{\text{ext1}}, F_{\text{ext2}} \in [-5,5]$, 
$\omega_1, \omega_2 \in [0,5].$
For 50 randomly selected combinations of force amplitudes and forcing frequencies within the specified ranges, the system of differential equations was integrated over a duration of ten seconds using the fourth-order Runge–Kutta method with a time-step of 0.01 sec. The resulting 50 time series of generalized coordinates and velocities, together with the corresponding time series of accelerations and forces, were assembled to form the dataset for the latent Lagrangian neural network corresponding to the 2-DOF system. Once the dataset was generated, the L-LNN was trained using 80\% of the dataset, following the procedure outlined in Algorithm~\ref{alg:llnn}, with the key hyperparameter values summarized in Table~\ref{tab:hyp}. Source code to reproduce the training data and the results will be provided on request.

On the other hand,  for the 5-DOF case the parameters of the model in appendix~\ref{app:5DOF} were set as $m = 1$, $k = 1$, $c = 0.9$, $\alpha_* = 0.5$ with initial conditions $q_i(0) = \dot{q}_i(0) = 0, \forall i = 1,\dots,5$, and external excitation applied at the fifth DOF as
$F_{\text{ext}}(t) = F_{\text{ext}} \sin(\omega t)$, 
$F_{\text{ext}} \in [-1,1], \quad \omega \in [0,5].$ For 70 randomly chosen combinations of force amplitudes and forcing frequencies within the specified ranges, the equations of motion were integrated for ten seconds using the fourth-order Runge–Kutta method with a time-step of 0.01 sec. The resulting time series of generalized coordinates, velocities from these 70 simulations and the corresponding accelerations and forces were then assembled to construct the dataset for learning the parameters of the latent Lagrangian neural network for the 5-DOF system. Following the generation of the dataset, the training of the L-LNN was carried out using 80\% of the dataset in accordance with Algorithm~\ref{alg:llnn}, and the corresponding values of the key hyperparameters are listed in Table~\ref{tab:hyp}. 

{It is important to note that the hyperparameter $\lambda$, present in Eq.~\eqref{eq:loss}, plays a critical role in controlling the relative weighting between reconstruction accuracy and the enforcement of the physics-based regularization within the model. In the present study, several values of $\lambda$ were explored for both systems under consideration.

For the first system, namely the 2DOF case, three different values of $\lambda$ ($\lambda = 1$, $0.05$, and $0.005$) were evaluated. The results indicated that neither the final prediction accuracy nor the convergence behavior exhibited significant sensitivity to the choice of $\lambda$ within this range. Consequently, the results reported for this system correspond to $\lambda = 1$, selected arbitrarily.

In contrast, similar investigations for the 5DOF system revealed a strong dependence on the choice of $\lambda$. Specifically, the value of this hyperparameter was found to significantly affect both the convergence rate and the robustness of the trained model with respect to model uncertainty, as assessed through repeated training with different random seeds. Among the tested values, $\lambda = 0.005$ consistently produced the most stable and reliable performance; therefore, the corresponding results are reported for the 5DOF case. 

Additionally, it is worth noting the specific constraints on the choice of activation functions in the proposed approach.The incorporation of the Euler-Lagrange equation into the loss function through force terms requires the computation of network Hessians, rendering the widely adopted ReLU activation function unsuitable for this approach, as they will always yield a zero Hessian. Conversely, the tanh and sigmoid activation functions would induce vanishing gradients. Consequently, we adopted the GELU activation function in this study. Nevertheless, both ELU and GELU activations exhibited similar outcomes in terms of training efficiency and prediction accuracy in the early explorations.}

\begin{table}[h!]
\caption{Values of Hyperparameters}
\centering
\begin{tabular}{ccc}
\hline
Hyperparameter & 2-DOF & 5-DOF \\ \hline
Initialization (PE) & Xavier (Uniform) \cite{pmlr-v9-glorot10a} & Xavier (Uniform) \\ 
Encoder: hidden depth and width & 1 and 64 &  1 and 64 \\ 
Decoder: hidden depth and width & 1 and 64 &  1 and 64 \\ 
PE, hidden depth and width& 3 and 128    & 3 and 128 \\ 
KE, hidden depth and width & 1 and 128 &  1 and 64   \\ 
Learning rate(initial) & 1E-04    &  1E-04    \\ 
Batch size &  256    & 400    \\ 
Scheduler &  ReduceLROnPlateau  & ReduceLROnPlateau \\ 
$\lambda$ & 1.0 & 0.005 \\
Activation function &  GeLU~\cite{hendrycks2016gaussian}   & GeLU~\cite{hendrycks2016gaussian}    \\ \hline
\end{tabular}
\label{tab:hyp}
\end{table}

\section{Performance Validation: Results and Discussion}

In this section, we present and discuss the main results obtained by applying the contribution presented in Section~\ref{sec:LLNN}, following the validation method described in Section~\ref{sec:demos}, to both test cases, namely the 2-DOF and 5-DOF nonlinear systems. For each system, the predicted response and the learned energy topology are compared against those obtained from the corresponding true vector field. {Further, to assess the robustness of the developed model with respect to uncertainty arising from variability in learned parameters and the stochastic nature of the training process, the variance in prediction errors was evaluated across multiple training runs using different random seeds for both test cases. 
In addition, to examine the robustness of the learned models to uncertainty due to noise in the data, models were trained using noisy datasets generated by introducing zero-mean Gaussian noise with a standard deviation equal to $5\%$ of the standard deviation of the original signals. The observations and implications from the aforementioned comparison and the analysis of robustness to uncertainty in model as well as data are detailed in the following subsections.}

\begin{figure}[h!]
    \centering
    \includegraphics[width=\textwidth]{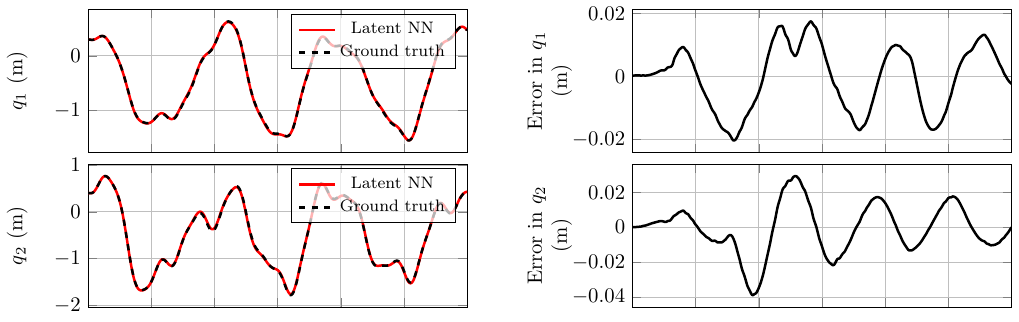}
    \caption{2-DOF system: displacements predicted through learned system vs the true displacements. The relative errors $e_{q_i}$ computed as per Algorithm~\ref{alg:validation} are 0.012 for $q_1$ and 0.016 for $q_2$.}
    \label{fig:2dof_disp}
\end{figure}
\begin{figure}[h!]
    \centering
    
    \includegraphics[width=\textwidth]{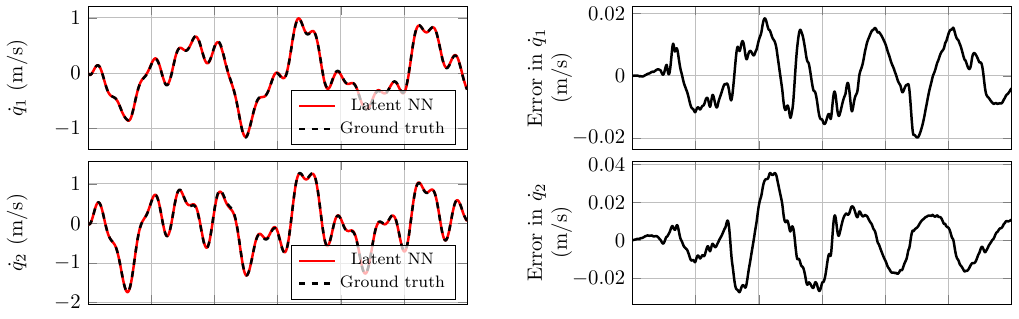}
    \caption{2-DOF system: velocities predicted through learned system vs the true velocities. The relative errors $e_{\dot{q}_i}$ computed as per Algorithm~\ref{alg:validation} is 0.019 for $\dot{q}_1$ and 0.020 for $\dot{q}_2$.}
    \label{fig:2dof_vel}
\end{figure}

\begin{figure}[h!]
    \centering

    \subfigure[$V_\textup{true}(q)$]{
        \includegraphics[width=0.47\textwidth]{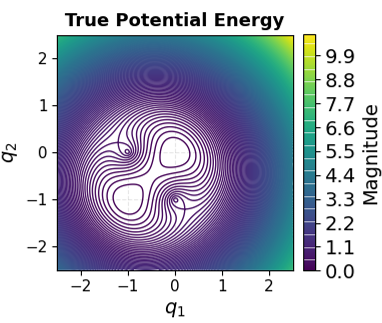}
    }
    \subfigure[$V_\textup{learned}(\mathcal{E}(q))$]{
        \includegraphics[width=0.47\textwidth]{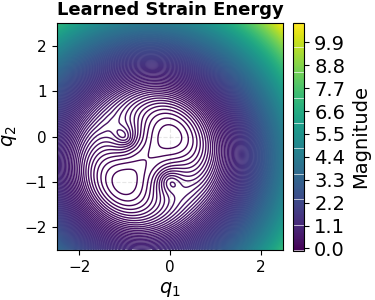}
    }

    \subfigure[$T_\textup{true}(\dot{q})$]{
        \includegraphics[width=0.47\textwidth]{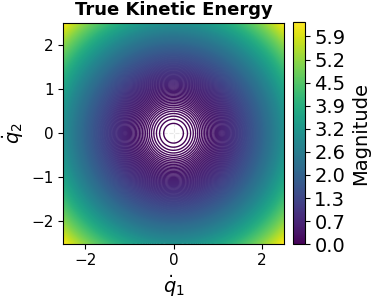}
    }
    \subfigure[$T_\textup{learned}(\mathcal{E}(q), J(q)\dot{q})$]{
        \includegraphics[width=0.47\textwidth]{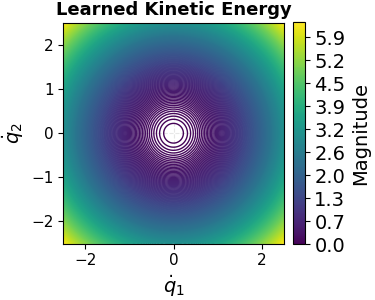}
    }
    \caption{Comparison of energy topology. The learned energies are converted in terms of original coordinates $q_1, q_2$ and $\dot q_1, \dot q_2$ to make a direct comparison of the contour lines relevant.}
    \label{fig:EnergyComparison}
\end{figure}

\subsection{2-DOF Nonlinear dynamical system with nonconvex potential energy}

The training data generated for the 2-DOF system (detailed in section~\ref{sec:test_cases})  by following the method mentioned in section~\ref{sec:data} has been used to implement Algorithm~\ref{alg:llnn}. This results in a set of three trained neural networks: an autoencoder, networks  representing potential and kinetic energy functions in the latent space of dimension two, which is equal to the original dimension of the configuration space in this test case. After obtaining the trained networks, their performance has been evaluated by obtaining the equations of motion (or the vector field) and  comparing the numerically calculated response (generalized coordinates, generalized velocity) from the learned system with the response obtained from the true system of equations, following the validation workflow presented in Algorithm~\ref{alg:validation}. 

The forcing functions used during validation are arbitrarily chosen as $F_1(t) = 0.5(\sin(2t) + \sin(4t))$  and  $F_2(t) = \sin(2t) + \sin(4t)$; the initial conditions are $q_1(0) = 0.3$, $q_2(0) = 0.4$  $\dot{q}_1(0) = \dot{q}_2(0) = 0$.  Note that, in contrast to the training data generation phase, the displacement initial conditions used during validation are deliberately chosen to be nonzero and nonsymmetric, and the applied forcing functions are non-harmonic. This setup is intended to demonstrate the ability of the method to capture the underlying physics even when the initial conditions and forcing functions differ from those utilized in the generation of training data. Figures~\ref{fig:2dof_disp} and \ref{fig:2dof_vel} show the results of the above-mentioned comparison between the responses obtained via learned models and the ground truth. Despite non-trivial responses, the relative errors as defined in Algorithm~\ref{alg:validation}, for the displacement components were below 2\%, and the velocity responses also registered a relative error of around 2\%. 
It must be noted that the model does not explicitly incorporate the time information during training, but still getting such close match in the time evolution of the response is encouraging. 
We attribute it to the principled enforcement of strong physical bias in the loss function, which validates the efficacy of the approach developed in this work. Another important observation is that, in the proposed methodology, the time step used during training is independent of the time step employed during inference from the reduced model. For instance, reproducing the results shown in Figures~\ref{fig:2dof_disp} and~\ref{fig:2dof_vel} with twice the step size produced indistinguishable plots. The relative errors obtained with time steps of $0.02$~s and $0.01$~s were identical when rounded to two decimal places. Source code to reproduce the figures has been provided as supplementary material. 

To further evaluate and interpret the learned system, the learned energy contours have been compared with the true energy contours. Because the system was chosen as a 2-DOF system, the energies are fully characterized by their contour plots (in 2D), offering an insightful comparison between ground truth and learned potential and kinetic energies, see Figure~\ref{fig:EnergyComparison}. The contours are obtained by plotting the energies with respect to their corresponding independent variables; for the kinetic energy, which depends on both the generalized coordinates and velocities, the coordinates are fixed at zero. It is evident from the results that the potential energy contours reproduce the multiple equilibrium wells of the nonlinear system, while the kinetic energy contour correctly reflects the isotropy of the inertia.
However, it must be noticed that even though the locations of the critical points were correctly captured and the shapes of the contours match quite well qualitatively, the exact values of the learned potential energy differ slightly from those of the true values. The reason for this may be that the critical points depend on the gradient of potential energy, which is explicitly a part of the loss function, while the potential energy function itself does not appear in the loss function. This is not worrisome in terms of response (as we have seen in the results) and the stiffness behaviour because the equations of motion depend only on the gradients of the energies, which are learned correctly because of their presence in the loss function. 

\begin{figure}[h!]
    \centering
    \includegraphics[height=5.5cm]{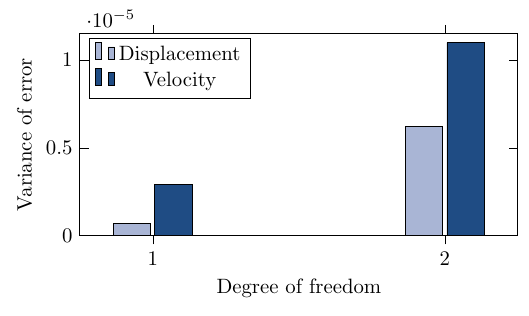}
    \caption{Variance of Relative Errors in Displacement and Velocity Responses (2-DOF System)}
    \label{fig:2dof_variance}
\end{figure}
The above-mentioned discussion on the learned kinetic and potential energy networks indicated that, unlike monolithic LNNs, which learn the Lagrangian as a single scalar function, the separation of potential and kinetic energy networks in L-LNN provides clear interpretability. It is worth mentioning that even though the parameters of the kinetic and potential energy networks are learned simultaneously using a unified loss function, the stiffness characteristics are encoded exclusively in the potential energy network, whereas inertia effects are captured by the kinetic energy network.  

Overall, the close agreement between learned and true energy contours and the response curves demonstrates that L-LNN not only predicts accurate trajectories but also recovers the governing energy structure of the system. This dual capability is crucial for extrapolation, stability analysis, and model interpretability in high-dimensional nonlinear dynamics.  {However, to further demonstrate the reliability of the obtained results and to evaluate the robustness of the proposed model with respect to model uncertainty, the latent-LNN was trained using three different random seeds, and the variance in the relative errors of the predicted displacement and velocity responses was computed. The results shown in Figure~\ref{fig:2dof_variance} indicate that the variability in the predicted responses across different random seeds was relatively small (i.e., of the order of $10^{-5}$), suggesting that the developed model exhibits stable performance with respect to stochastic variations in the training process.}

\begin{figure}[h!]
    \centering
    \includegraphics[width=\textwidth]{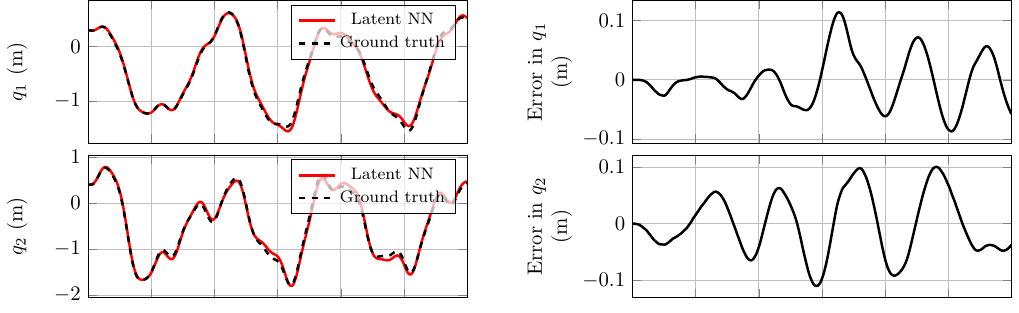}
    \caption{2-DOF system: displacements predicted through system trained with noisy data vs the true displacements. The relative errors $e_{q_i}$ computed as per Algorithm~\ref{alg:validation} is 0.051 for $q_1$ and 0.062 for $q_2$.}
    \label{fig:2dof_disp_noisy}
\end{figure}
\begin{figure}[h!]
    \centering
    
    \includegraphics[width=\textwidth]{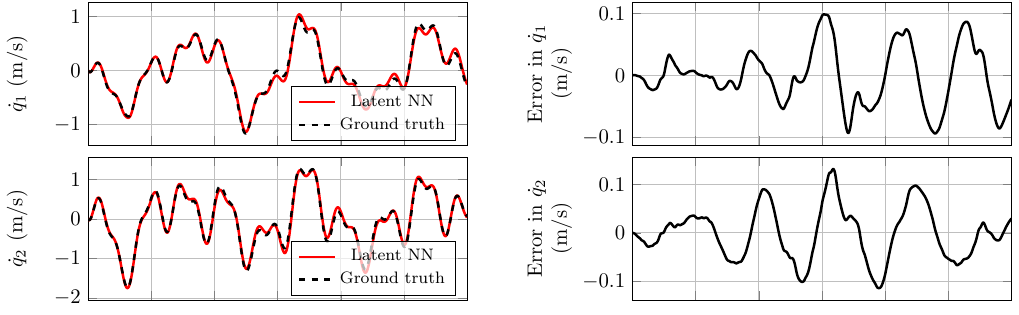}
    \caption{2-DOF system: velocities predicted through system trained with noisy data vs the true velocities. The relative errors $e_{\dot{q}_i}$ computed as per Algorithm~\ref{alg:validation} are 0.097 for $\dot{q}_1$ and 0.083 for $\dot{q}_2$.}
    \label{fig:2dof_vel_noisy}
\end{figure}

{Furthermore, to assess the robustness of the model to noisy inputs and to evaluate its direct applicability to experimental scenarios, additional models were trained using noisy datasets generated by introducing zero-mean Gaussian noise with a standard deviation equal to $5\%$ of the standard deviation of the original signals. The corresponding responses are presented to illustrate the effect of data noise on the prediction error. Although a moderate increase in the relative errors is observed, the overall agreement between the learned and reference responses remains satisfactory, indicating that the developed model maintains reliable predictive capability in the presence of noise in the data. Notably, an alternative strategy in practical applications could include a data preprocessing step to reduce noise prior to model training; however, directly training the model using noisy datasets provides a more stringent assessment of the robustness of the proposed framework.}

The 2-DOF case offered the possibility to directly compare the level sets of the energies in 2D. However, for this precise reason, it does not give insight into model reduction, since the latent dimension equals the original dimension. This shortcoming is overcome in the next section with validation on a  higher-dimensional system.

\subsection{5-DOF Nonlinear dynamical system with cubic nonlinearity}

\begin{figure}[h!]
    \centering
    \includegraphics[width=\textwidth]{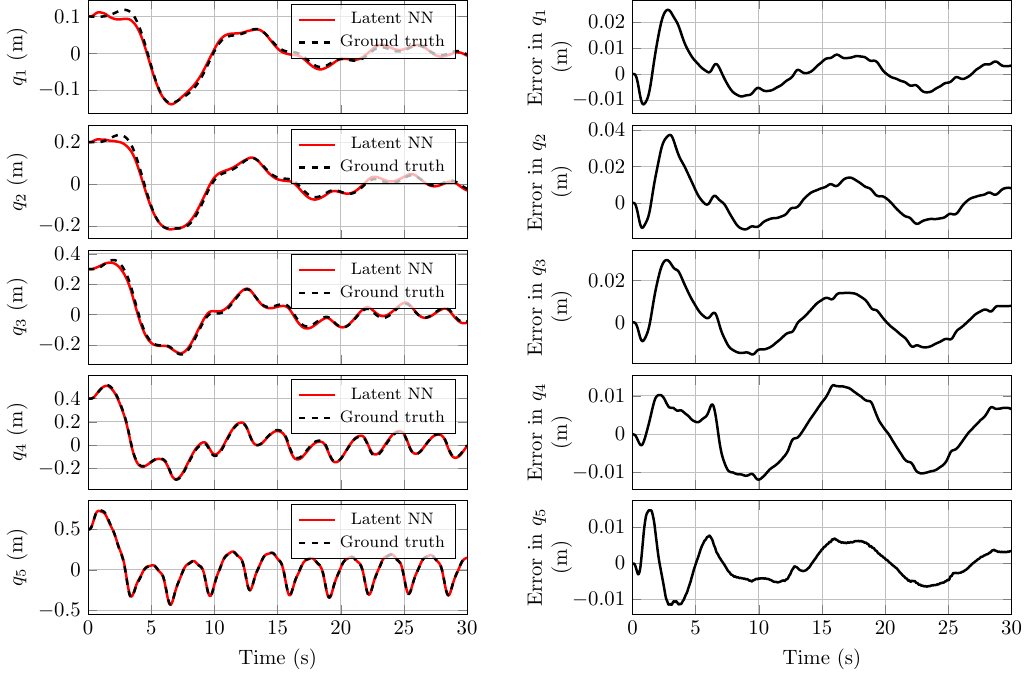}
   \caption{5-DOF system: displacements predicted through learned system  vs true displacements. The relative error $e_{q_i}$ computed as per Algorithm~\ref{alg:validation} are respectively 0.12, 0.10, 0.079, 0.043 and 0.023, for $i$ from 1 to 5.}
   \label{fig:disp5dof}
\end{figure}

\begin{figure}[h!]
    \centering
    \includegraphics[width=\textwidth]{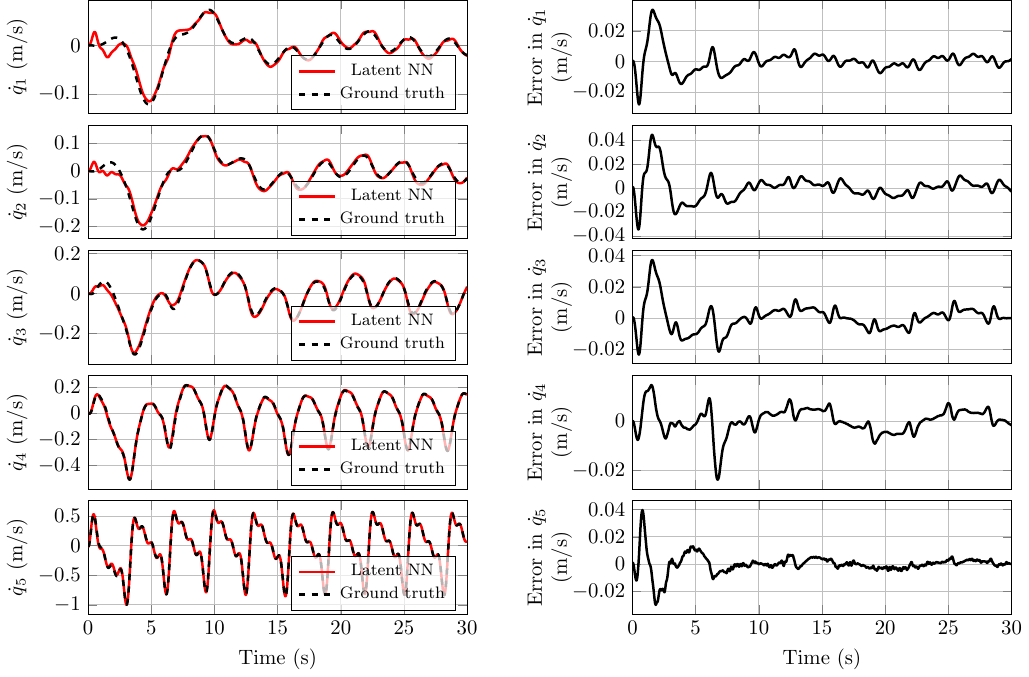}
    \caption{5-DOF system: velocities predicted through learned system  vs true velocities. The relative errors $e_{\dot{q}_i}$ computed as per Algorithm~\ref{alg:validation} are respectively 0.21, 0.16, 0.094, 0.033 and 0.018, for $i$ from 1 to 5.} 
    \label{fig:vel5dof}
\end{figure}

\begin{figure}[h!]
    \centering

    \subfigure[$T_\textup{true}(\dot{q})$]{
        \includegraphics[width=0.45\textwidth]{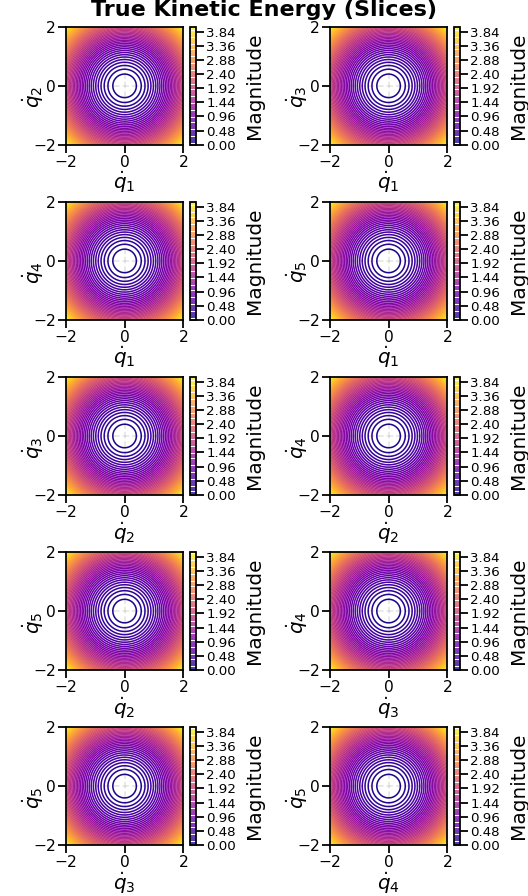}
    }
    \subfigure[$T_\textup{learned}(\mathcal{E}(q), J(q)\dot{q})$]{
        \includegraphics[width=0.45\textwidth]{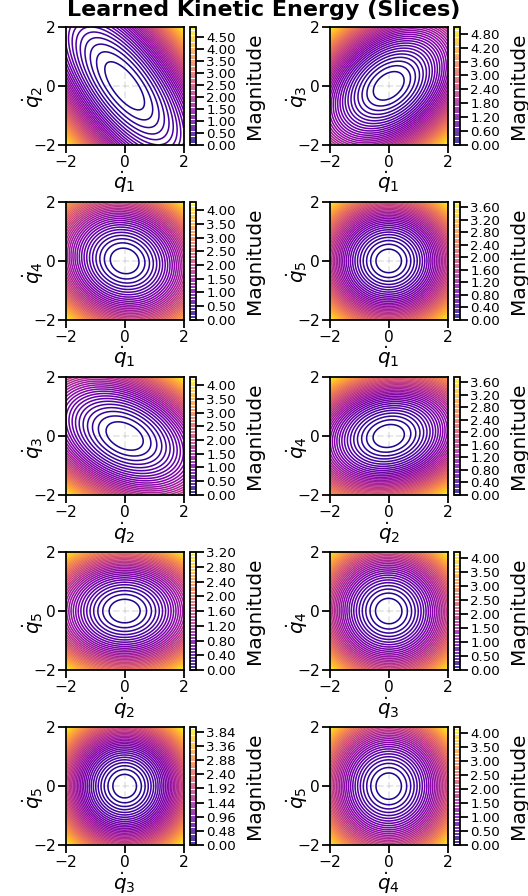}
    }
    
    \caption{5-DOF system: Kinetic Energy Comparison} 
    \label{fig:kin_cont_5dof}
\end{figure}

\begin{figure}[h!]
    \centering

    \subfigure[$V_\textup{true}(q)$]{
        \includegraphics[width=0.45\textwidth]{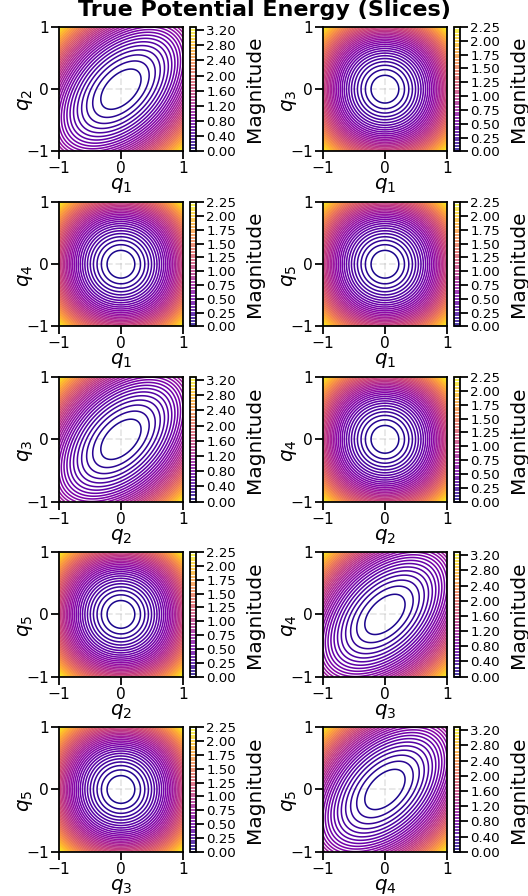}
    }
    \subfigure[$V_\textup{learned}(\mathcal{E}(q))$]{
        \includegraphics[width=0.45\textwidth]{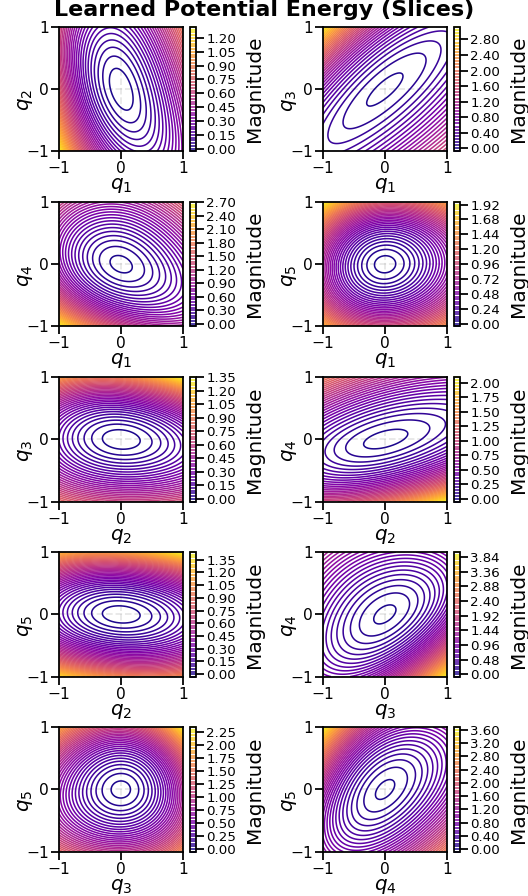}
    }
    
    \caption{5-DOF system: Potential Energy Comparison }
    \label{fig:pot_cont_5dof}
\end{figure}
This subsection will further strengthen the validity of the developed approach by implementing it for the 5-DOF system shown in Figure~\ref{fig:5dof} and detailed in section~\ref{sec:test_cases}. The training data generated for this system, as described in Section~\ref{sec:data}, has been used to implement Algorithm~\ref{alg:llnn}. As before, this implementation yields three jointly-trained neural networks: an autoencoder and two networks representing the potential and kinetic energy functions in the latent space. For this system, the latent space dimension was chosen to be 4, representing a reduction from the original 5-dimensional configuration space. Further reduction in dimension led to higher relative errors. After obtaining the trained networks, their performance has been evaluated by obtaining the equations of motion (or the vector field) and  comparing the numerically calculated response (generalized coordinates, generalized velocity) from the learned system with that of the true system of equations, following the validation workflow presented in Algorithm~\ref{alg:validation}. 

\begin{figure}[h!]
    \centering
    \includegraphics[height=5.5cm]{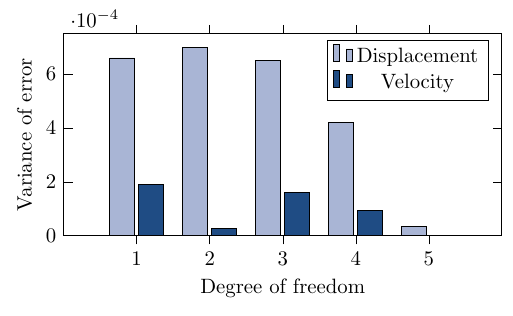}
    \caption{Variance of Relative Errors in Displacement and Velocity Responses (5-DOF System)}
    \label{fig:5dof_variance}
\end{figure}

The forcing functions used during validation is $F(t) = \begin{bmatrix}
0 & 0 & 0 & 0 & F_5(t)
\end{bmatrix}^T$
where $F_5(t)= \sum_{n=1}^{4} \sin(2nt)$ and the initial conditions are $q_1(0) = 0.1$, $q_2(0) = 0.2$, $q_3(0) = 0.3$, $q_4(0) = 0.4$, $q_5(0) = 0.5$ and  $v_1(0) = \dots = v_5(0) = 0$. Note that, again, in contrast to the training data generation, validation uses nonzero, nonsymmetric initial displacements and non-harmonic forcing, thereby testing the method’s ability to capture the underlying physics under conditions distinct from those used in training.

Figures~\ref{fig:disp5dof} and \ref{fig:vel5dof} show the results of the above-mentioned comparison between the learned response with that of the ground truth. The displacement responses predicted via the learned vector field exhibit excellent agreement 
with the true system responses across all five degrees of freedom. The relative errors 
for each of the displacement components remain small, with 
$e_{q_1} = 12.1\%$, 
$e_{q_2} = 10.5\%$, 
$e_{q_3} = 7.9\%$, 
$e_{q_4} = 4.3\%$, 
and $e_{q_5} = 2.3\%$. 
The predicted displacement responses accurately capture both amplitude and phase characteristics 
of the system motion. Similarly, for the velocity responses, the neural predictions closely follow the ground-truth. The relative errors are $21\%$ and $16.4\%$ for the relatively slower components $\dot{q}_1$ and $\dot{q}_2$ respectively, while they remain below $10\%$ for the other velocity components. In terms of time evolution of the system, the overall match is very good, demonstrating the model’s capability to generalize well across all states of the 5-DOF nonlinear system. Such a close agreement between the predicted and the true displacement and velocity functions substantiates the validity of the developed methodology which stems from its ability to maintain the virtual work consistency during the projection of latent force vectors to the original space, and the enforcement of bias from Euler-Lagrange equation in the loss function even when the latent space dimension is smaller than the original dimension of the configuration space.
\begin{figure}[h!]
    \centering
    \includegraphics[width=\textwidth]{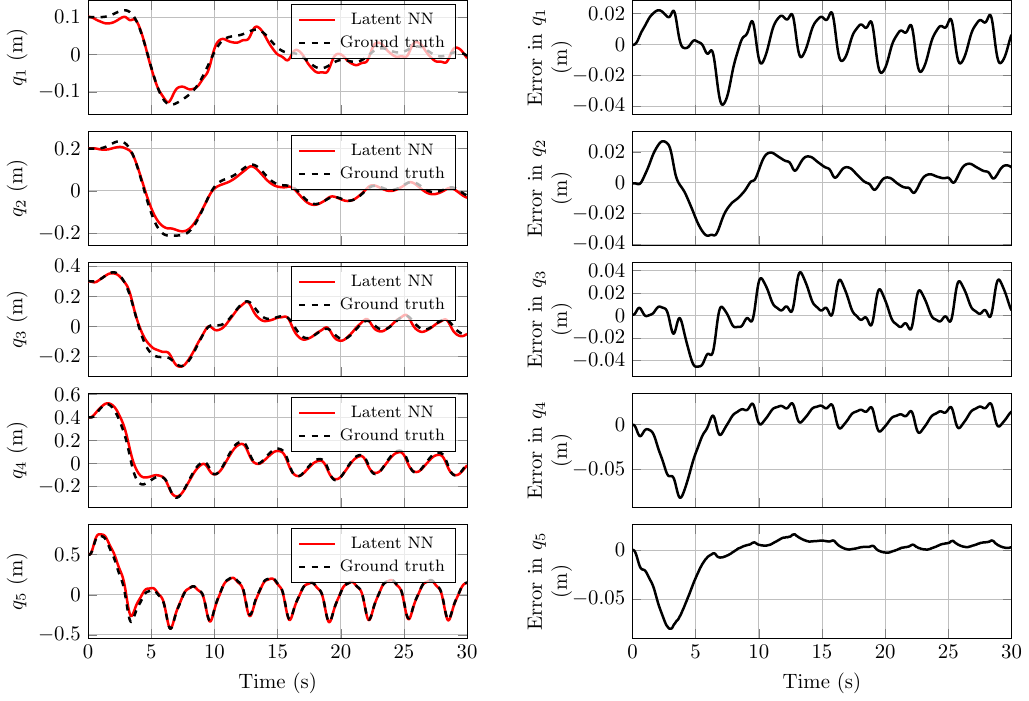}

   \caption{5-DOF system: displacements predicted through system trained with noisy data vs the true displacements. The relative errors $e_{q_i}$ computed as per Algorithm~\ref{alg:validation} are 0.22, 0.12, 0.12, 0.13 and 0.096, for $i$ from 1 to 5.} 
   \label{fig:noisy_disp5dof}
\end{figure}

\begin{figure}[h!]
    \centering
    \includegraphics[width=\textwidth]{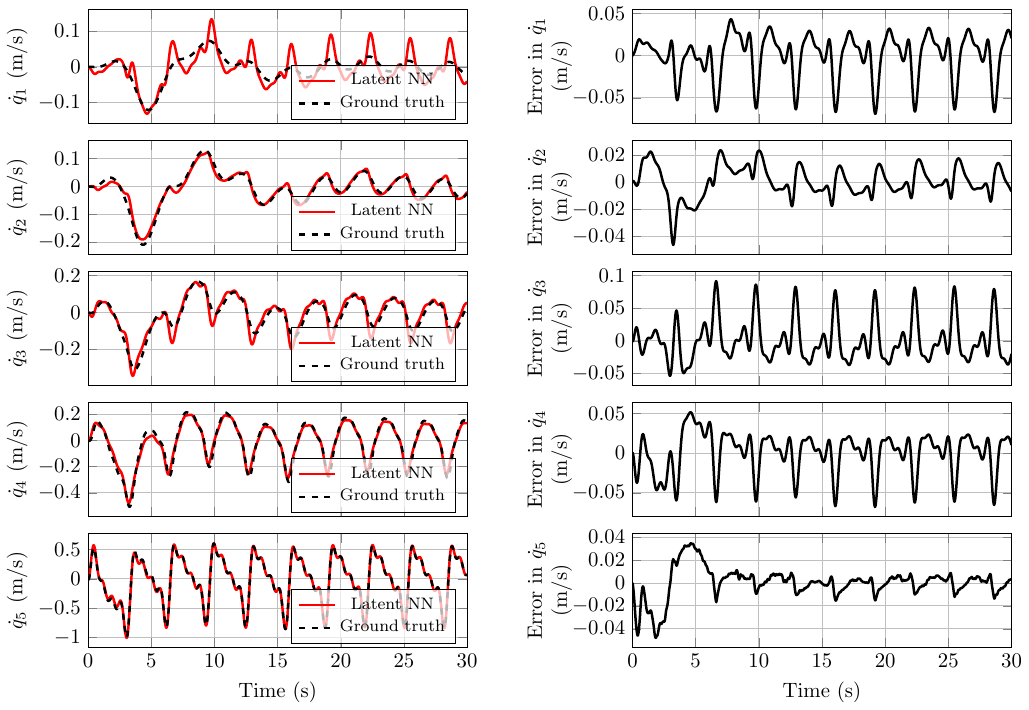}
    \caption{5-DOF system: velocities predicted through system trained with noisy data vs the true velocities. The relative errors $e_{\dot{q}_i}$ computed as per Algorithm~\ref{alg:validation} are 0.73, 0.18, 0.34, 0.16, 0.035, for $i$ for 1 to 5.}
    \caption{Comparison of Velocity Results for the Noisy 5-DOF System} 
    \label{fig:noisy_vel5dof}
\end{figure}

Although the complete energy surface in the 5D configuration space cannot be visualized, we compare the 2D slices by keeping the other dimensions fixed at zero for both the learned potential and Kinetic energy with the true ones, as shown in Figure~\ref{fig:kin_cont_5dof} and Figure~\ref{fig:pot_cont_5dof}. 

As expected, the contours did not match exactly for the 5-DOF system due to the difference in the latent dimension and the original dimension of the configuration space. For this case, the circular contours of the true kinetic energy do not retain their exact structure for the learned kinetic energy function due to the inevitable cross-coupling terms and redistribution of kinetic energy associated with different velocity components that arise due to the encoding of five dimensions in a four-dimensional latent space. However, the learned kinetic energy approximates the scale of the true kinetic energy well, and the general quadratic form and convex nature of the kinetic energy are well preserved. Similarly, in the case of potential energy, the distortions and rotations visible in the learned potential energy function with respect to the true potential energy can be attributed to the encoding of five coordinates into four latent coordinates. Thus, the observed deviations are not simply the artefacts of training but are inevitable when the reduction of dimensionality of the system is involved. However, it must be noticed that the deviations in potential energy are slightly more than the kinetic energy; this might be due to the fact that the kinetic energy network possesses a strong bias in its architecture, which is enforced to be quadratic in the latent velocity.
 
 {Similar to the first test case of the 2DOF system, to assess the reliability of the obtained results and the robustness of the proposed model with respect to model uncertainty, the latent-LNN has been trained using three different random seeds, and the variance in the relative errors of the predicted displacement and velocity responses have been evaluated. The result shown in Figure~\ref{fig:5dof_variance} indicate that the variability across different random seeds is quite small (i.e., of the order of $10^{-4}$), suggesting stable model performance with respect to stochastic variations in the training process.}

{Furthermore, to examine robustness to noisy inputs and evaluate direct applicability to experimental scenarios, additional models were trained using datasets mixed with zero-mean Gaussian noise having a standard deviation equal to $5\%$ of that of the original signals.  It should be noted that directly adding noise to the same dataset used in the noise-free case did not yield satisfactory accuracy. This was primarily because the responses of the first and second masses (from the left fixed end) had magnitudes comparable to the noise level added to the response of the fifth mass, where the external forcing was applied during data generation. As a result, the signal-to-noise ratio for the lower-amplitude responses became unfavorable, which adversely affected the training accuracy. To mitigate this issue, the noise was added to a dataset generated by exciting the second mass as well in addition to the fifth mass with $\omega \in [0,12]$, thereby increasing the richness of the training data and enabling more accurate learning of the system dynamics. Furthermore, the number of response samples used for training in the noisy case corresponds to 80$\%$ of 100 samples, as opposed to 80$\%$ of 70 samples in the noise-free case. The displacement and velocity responses obtained from the system trained using the noisy dataset and the corresponding relative errors are presented in figures~\ref{fig:noisy_disp5dof} and \ref{fig:noisy_vel5dof}. These figures clearly demonstrate the effect of data noise on prediction errors. Although a noticeable increase in relative errors is observed compared to the noise-free case, the overall agreement between the predicted and true responses remains satisfactory, depicting predictive capability in the presence of noise. While noise reduction through data preprocessing can be employed in practice to improve accuracy, directly training on noisy datasets for this validation study provided a more stringent assessment of the robustness of the proposed framework.}
 
Hence, it can be inferred from the discussion of the results presented in this subsection that the developed approach has the ability to closely estimate the original trajectory even when the energies (Lagrangian) are learned in a latent space of dimension lower than the original dimension of the configuration space. In addition, the modular approach for learning the Lagrangian by decomposing it into separate potential and kinetic energy networks not only enhances the control over the network architectures but also improves interpretability. 
 
{Finally, it is worth mentioning that the obtained results demonstrate highly accurate prediction of displacement and velocity responses for both test cases. However, it is important to note that precise long-term response prediction in chaotic systems is fundamentally limited due to their sensitive dependence on initial conditions. In such systems, small discrepancies in the initial state grow exponentially over time, leading to divergence of trajectories even for well-trained models. Consequently, the primary objective in chaotic dynamics is not the exact prediction of individual trajectories but rather the faithful reproduction of the system’s long-term statistical and dynamical properties~\cite{nakamura2021convolutional, linot2020deep}.}

\section{Conclusion}

A new physically grounded framework for the reduced-order modelling of high-dimensional nonlinear dynamical systems has been developed. The method is based on learning the Lagrangian of a dynamical system in a latent space of smaller dimension by decomposing it into three neural networks:  one autoencoder for the generalized coordinates,  one for learning the potential energy, and another one for the kinetic energy. This can be seen either as an extension of Lagrangian NNs to model reduction, or as an extension of Latent-Energy-based NNs to dynamics. 
The parameters of each of these networks have been learned simultaneously using a common loss function derived from the Euler-Lagrange Equations and the virtual work principle. The efficacy of the developed framework has been validated by demonstrating its performance in predicting the behaviour of two different nonlinear dynamical systems. It showed a remarkable predictive performance via training with less than 60 time series of 10 seconds duration each for both the test cases. Moreover, some of its salient features, such as the modular control over the network architectures for the representation of the potential and the virtual work consistency between the true and the latent force fields, provided us not just the better control over their architectures but an improved interpretability of the learned energies as well. The encouraging results obtained for the prototypical problems will motivate their future application to physics-informed data-driven reduced order modelling of more challenging nonlinear mechanical systems in the research community. 

\section*{Acknowledgements}
The authors would like to thank Louen Pottier for helpful discussions on LLNN and the choice of test cases, which contributed to
this work. The authors also thank the reviewers for their constructive comments, which helped improve the manuscript.

\section*{References}
\printbibliography[heading=bibempty]

\appendix

\section{Equations of motion for the test cases}

\subsection{2-DOF nonlinear spring-mass system}
\label{app:2DOF}
The 2-DOF spring-mass system can be mathematically described by the generalized coordinates $q_1(t)$ and $q_2(t)$, thus, the equations of motions can be given as follows: 
\begin{align}
    m \ddot{q}_1 + \frac{\partial V}{\partial q_1} + c_x \dot{q}_1 
    &= F_1(t), \\[0.5em]
    m \ddot{q}_2 + \frac{\partial V}{\partial q_2} + c_y \dot{q}_2 
    &= F_2(t).
\end{align}
The initial conditions are specified as
\begin{align}
q_1(0) = q_{1,0}, \quad \dot{q}_1(0) = v_{1,0}, \\[0.5em]
q_2(0) = q_{2,0}, \quad \dot{q}_2(0) = v_{2,0}.
\end{align}
where the potential energy of the system is given as:
\begin{equation}
V(q_1,q_2) = \tfrac{1}{2} k_x \big( L_x - L_{0x} \big)^2 
           + \tfrac{1}{2} k_y \big( L_y - L_{0y} \big)^2,
\end{equation}
such that
\begin{align}
L_x &= \sqrt{(L_{0x}+q_1)^2 + q_2^2}, \\[0.5em]
L_y &= \sqrt{q_1^2 + (L_{0y}+q_2)^2}.
\end{align}
and the kinetic energy is
\begin{equation}
T = \tfrac{1}{2} m \left( \dot{q}_1^2 + \dot{q}_2^2 \right)
= \frac{1}{2} m \dot{q}_1^2 + \frac{1}{2} m \dot{q}_2^2.
\end{equation}
the rest of the parameters can be described as follows: 
\begin{itemize}
  \item $m$: mass of the particle.  
  \item $k_x, k_y$: spring stiffness constants in the $x$ and $y$ directions.  
  \item $L_{0x}, L_{0y}$: \emph{free lengths} of the springs in the $x$ and $y$ directions (their natural lengths without load).  
  \item $c_x, c_y$: viscous damping coefficients in the $x$ and $y$ directions.
  \item $F_1(t)$ and $F_2(t)$ are external forcing functions which are chosen to be harmonic for training as: $F_{\text{ext1}} \sin(\omega_1 t)$ and $F_{\text{ext2}} \sin(\omega_2 t).$
  \item $F_{\text{ext1}}, F_{\text{ext2}}$: amplitudes of the external harmonic forces in the $x$ and $y$ directions.  
  \item $\omega_1, \omega_2$: forcing frequencies corresponding to $F_{\text{ext1}}$ and $F_{\text{ext2}}$.  
  \item $q_{1,0}, q_{2,0}$: initial displacements.  
  \item $v_{1,0}, v_{2,0}$: initial velocities.  
\end{itemize}

\subsection{5-DOF nonlinear spring–mass system}
\label{app:5DOF}
If we denote the generalized coordinates as $q_1(t)$, $q_2(t)$, $q_3(t)$, $q_4(t)$, and $q_5(t)$,  
the 5-DOF system can be mathematically described by the following matrix equation:

\begin{equation}
M \ddot{q}(t) + C \dot{q}(t) + K_{\text{eff}}(q)\, q(t) = F(t),
\end{equation}

The initial conditions for this system are specified as:
\begin{equation}
q(0) = q_0, \qquad \dot{q}(0) = v_0.
\end{equation}

where the effective stiffness matrix is
\begin{equation}
K_{\text{eff}}(q) = K + K_{\text{nl}}(q).
\end{equation}

The equations for kinetic and potential energy of this system can be described as:
The kinetic energy is
\begin{equation}
T = \tfrac{1}{2}\,\dot{q}^T M \dot{q} 
   = \frac{m}{2}\left(\dot{q}_1^2 + \dot{q}_2^2 + \dot{q}_3^2 + \dot{q}_4^2 + \dot{q}_5^2\right).
\label{eq:KE}
\end{equation}
and the expression for potential energy is
\begin{equation}
V = \tfrac{1}{2}\,q^T K q + \frac{\alpha_*}{4}\left(q_1^4 + q_2^4 + q_3^4 + q_4^4 + q_5^4\right).
\end{equation}

The system parameters are described below: 

\begin{itemize}
  \item Displacement/Generalized coordinate vector: 
\[
q(t) = 
\begin{bmatrix}
q_1(t) & q_2(t) & q_3(t) & q_4(t) & q_5(t)
\end{bmatrix}^T.
\]

  \item Mass matrix: 
  \[
  M = m
  \begin{bmatrix}
  1 & 0 & 0 & 0 & 0 \\
  0 & 1 & 0 & 0 & 0 \\
  0 & 0 & 1 & 0 & 0 \\
  0 & 0 & 0 & 1 & 0 \\
  0 & 0 & 0 & 0 & 1
  \end{bmatrix}.
  \]

  \item Damping matrix:
  \[
  C = c
  \begin{bmatrix}
  2 & -1 & 0 & 0 & 0 \\
  -1 & 2 & -1 & 0 & 0 \\
  0 & -1 & 2 & -1 & 0 \\
  0 & 0 & -1 & 2 & -1 \\
  0 & 0 & 0 & -1 & 2
  \end{bmatrix}.
  \]

  \item Linear stiffness matrix:
  \[
  K = k
  \begin{bmatrix}
  2 & -1 & 0 & 0 & 0 \\
  -1 & 2 & -1 & 0 & 0 \\
  0 & -1 & 2 & -1 & 0 \\
  0 & 0 & -1 & 2 & -1 \\
  0 & 0 & 0 & -1 & 2
  \end{bmatrix}.
  \]

  \item Nonlinear stiffness matrix:
  \[
  K_{\text{nl}}(q) = \alpha_* 
  \begin{bmatrix}
  q_1^2 & 0 & 0 & 0 & 0 \\
  0 & q_2^2 & 0 & 0 & 0 \\
  0 & 0 & q_3^2 & 0 & 0 \\
  0 & 0 & 0 & q_4^2 & 0 \\
  0 & 0 & 0 & 0 & q_5^2
  \end{bmatrix}.
  \]
  
here $\alpha_*$ is the nonlinear stiffness coefficient that controls the strength 
of the cubic restoring force contribution.
\item External force vector:
\[
F(t) = 
\begin{bmatrix}
0 & 0 & 0 & 0 & F_{\text{ext}} \sin(\omega t)
\end{bmatrix}^T.
\]
\end{itemize}

\end{document}